\documentclass[aps,prb,twocolumn,showpacs,psfig,longbibliography]{revtex4-2}

\usepackage[compat=1.0.0]{tikz-feynman}
\usepackage{amssymb}
\usepackage{multirow}
\usepackage{bm}
\usepackage{amsmath}
\usepackage{graphicx}
\usepackage{epstopdf}
\usepackage{subfigure}
\usepackage{natbib}
\usepackage{epsfig}
\usepackage{amsfonts}
\usepackage{mathrsfs}
\usepackage{braket}
\usepackage[toc,page,title,titletoc,header]{appendix}
\usepackage[colorlinks,linkcolor=blue,citecolor=blue]{hyperref}
\usepackage{dsfont,amsthm,amsbsy}

\usetikzlibrary{quotes,angles}
\newcommand{\obtuseangle}{\kern.08em
\begin{tikzpicture}
    \draw coordinate (a) at (0.14,0);
    \draw coordinate (b) at (0,0);
    \draw coordinate (c) at (-.12,0.18);
    \draw (a) -- (b) -- (c) pic [draw=black]{} ;
\end{tikzpicture}%
\kern.08em%
}

\begin{document}
\title{Pair density wave in doped three-band Hubbard model on two-leg square cylinders}
\author{Hong-Chen Jiang}
\email{hcjiang@stanford.edu}
\affiliation{Stanford Institute for Materials and Energy Sciences, SLAC National Accelerator Laboratory and Stanford University, Menlo Park, California 94025, USA}

\date{\today}
\begin{abstract}
A pair density wave (PDW) is a superconducting (SC) state with spatially modulated order parameter. Although much is known about the properties of the PDW state, its realization in microscopic models with divergent susceptibility has been challenging. Here we report a density-matrix renormalization group study of a three-band Hubbard model (also known as the Emery model) for cuprates on long two-leg square cylinders. Upon light doping, we find that the ground state of the system is consistent with that of a PDW state with mutually commensurate and power-law SC, charge (CDW) and spin (SDW) density wave correlations. The SC correlations are dominant between neighboring Cu sites with d-wave pairing symmetry. Interestingly, we find that the near-neighbor interactions, especially the near-neighbor attractive $V_{pd}$ interaction between neighboring Cu and oxygen sites, can notably enhance the SC correlations while simultaneously suppressing the CDW correlations. For a modestly strong attractive $V_{pd}$, the SC correlations become quasi-long-ranged with a divergent PDW susceptibility.
\end{abstract}
\maketitle

\section{Introduction}%
The pair density wave (PDW) is a novel superconducting (SC) state in which the Cooper pairs carry finite center-of-mass momentum and the SC order parameter varies periodically in space in such a way that its spatial average vanishes \cite{Berg2009,Lee2014,Fradkin2015,Agterberg2020}. 
The first example of PDW is the Fulde-Ferrell-Larkin-Ovchinnikov state in which the PDW state is more stable than the spatially uniform SC state when the Fermi surfaces are split by an external magnetic field \cite{FF1964,LO1965}. The PDW state has been discussed as a leading candidate state to understand the physics of cuprate high-temperature superconductors and other strongly correlated systems, where it has been proposed that various phases, including the unconventional superconductivity, charge density wave (CDW) and spin density wave (SDW) orders, can emerge by partially melting the PDW state \cite{Lee2014,Fradkin2015,Agterberg2020,Himeda2002}. Recently, intense interest in the PDW state has emerged due to experimental observations in underdoped cuprate superconductors, where signatures of PDW states have been observed via local Cooper pair tunneling and scanning tunneling microscopy in underdoped ${\rm Bi_2Sr_2CaCu_2O_{8+x}}$ (BSCCO) \cite{Hamidian2016,Ruan2018,Edkins2019,Liu2021} and the dynamical inter-layer decoupling in La$_{1.875}$Ba$_{0.125}$CuO$_4$ (LBCO) \cite{Li2007,Agterberg2008,Berg2007,Tranquada2008,Tranquada2020,Tranquada2021,Lozano2021}. Signatures of PDW states have also been reported in kagome superconductor CsV$_3$Sb$_5$ \cite{Chen2021k} and monolayer iron-based high-Tc Fe(Te,Se) films \cite{Liu2022}.

While much is known about the properties of the PDW state \cite{Berg2009,Lee2014,Fradkin2015,Agterberg2020,Tranquada2020,Tranquada2021,Wu2022}, there are very few microscopic models which are shown to have PDW ground states \cite{Berg2010,Fradkin2012,Venderley2019,Xu2019,Peng2021a,Peng2021b,Fradkin2012,Han2020,Huang2022} and especially no PDW long-range order has been established in unbiased calculations in two or higher dimensions. These include the one-dimensional (1D) Kondo-Heisenberg model \cite{Berg2010}, the extended Hubbard-Heisenberg model on two-leg ladder \cite{Fradkin2012}, generalized $t$-$J$ and Hubbard models \cite{Xu2019,Venderley2019,Peng2021a,Peng2021b} and strong coupling limit of Holstein-Hubbard model \cite{Han2020,Huang2022}. However, even in quasi-1D systems such as ladders, no reported evidence of PDW quasi-long-range order with divergent susceptibility -- the closest one can have to PDW long-range order -- in standard Hubbard model on any systems including two-leg ladder, especially that are closely related to real materials.

In the present work, we consider the three-band Hubbard model (also known as the Emery model) on the square lattice, which has long been proposed as one of the minimal models to describe the properties of cuprate high-temperature superconductors \cite{Zaanen1985,Emery1987,Zhang1988,White2015,Huang2017}, with the goal of determining whether this model supports PDW order. We employ density-matrix renormalization group (DMRG) \cite{White1992} to study the lightly hole doped three-band Hubbard model \cite{White2015} on long two-leg square cylinders. Our main results are that the ground state of the system is consistent with that of a PDW state with power-law and mutually commensurate SC, CDW and SDW correlations. The SC correlations oscillate periodically and change sign in real space. The PDW ordering wavevector $Q\approx 2\pi\delta$ is incommensurate and hole doping density $\delta$ dependent. The Cooper pairing is dominant between neighboring Cu sites with $d$-wave symmetry. Interestingly, we find that near-neighbor attractions, especially $V_{pd}$ between neighboring Cu and O sites, can notably enhance the SC correlations while suppressing the CDW correlations. For modestly strong near-neighbor attractions, the SC correlations become quasi-long-ranged with a divergent PDW susceptibility.

\begin{figure}
  \includegraphics[width=0.95\linewidth]{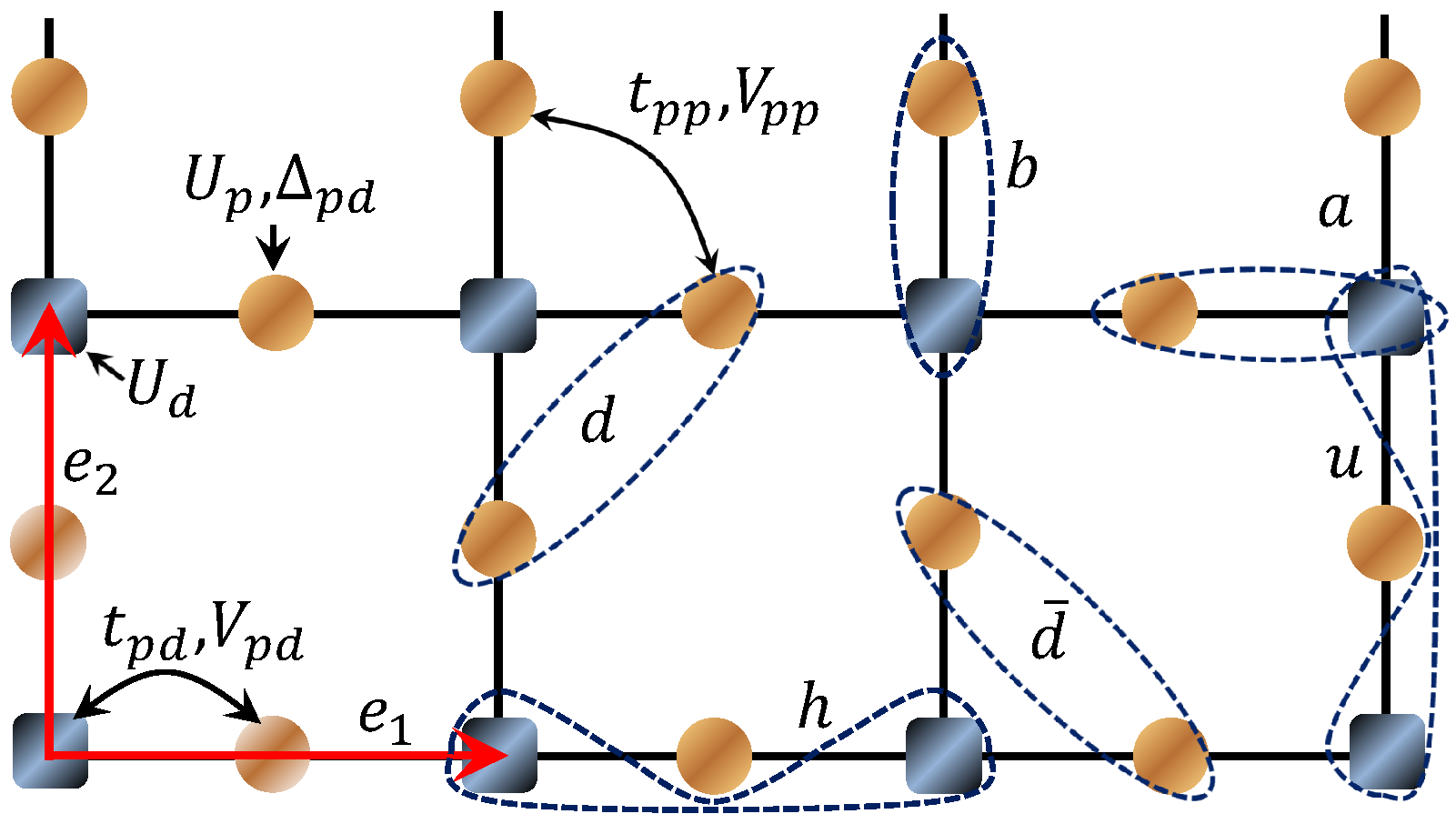}
\caption{Three-band Hubbard model on square lattice. The squares represent Cu $d_{x^2-y^2}$ orbitals and circles represent O 2$p_x$/$p_y$ orbitals. $U_d$ and $U_p$ are onsite Cu and O Coulomb interactions, $V_{pd}$ and $V_{pp}$ are near-neighbor Cu-O and O-O Coulomb interactions, $t_{pd}$ and $t_{pp}$ are near-neighbor Cu-O and O-O hole hopping matrix elements. $\Delta_{pd}$ is the energy difference between having a hole on the O versus Cu sites. $e_1=(1,0)$ and $e_2=(0,1)$ are lattice basis vectors. The dashed loops represent bonds $a$, $b$, $d$, $\bar{d}$, $h$ and $u$.} \label{Fig:Lattice} 
\end{figure}

\begin{figure*}
\centering
  \includegraphics[width=0.9\linewidth]{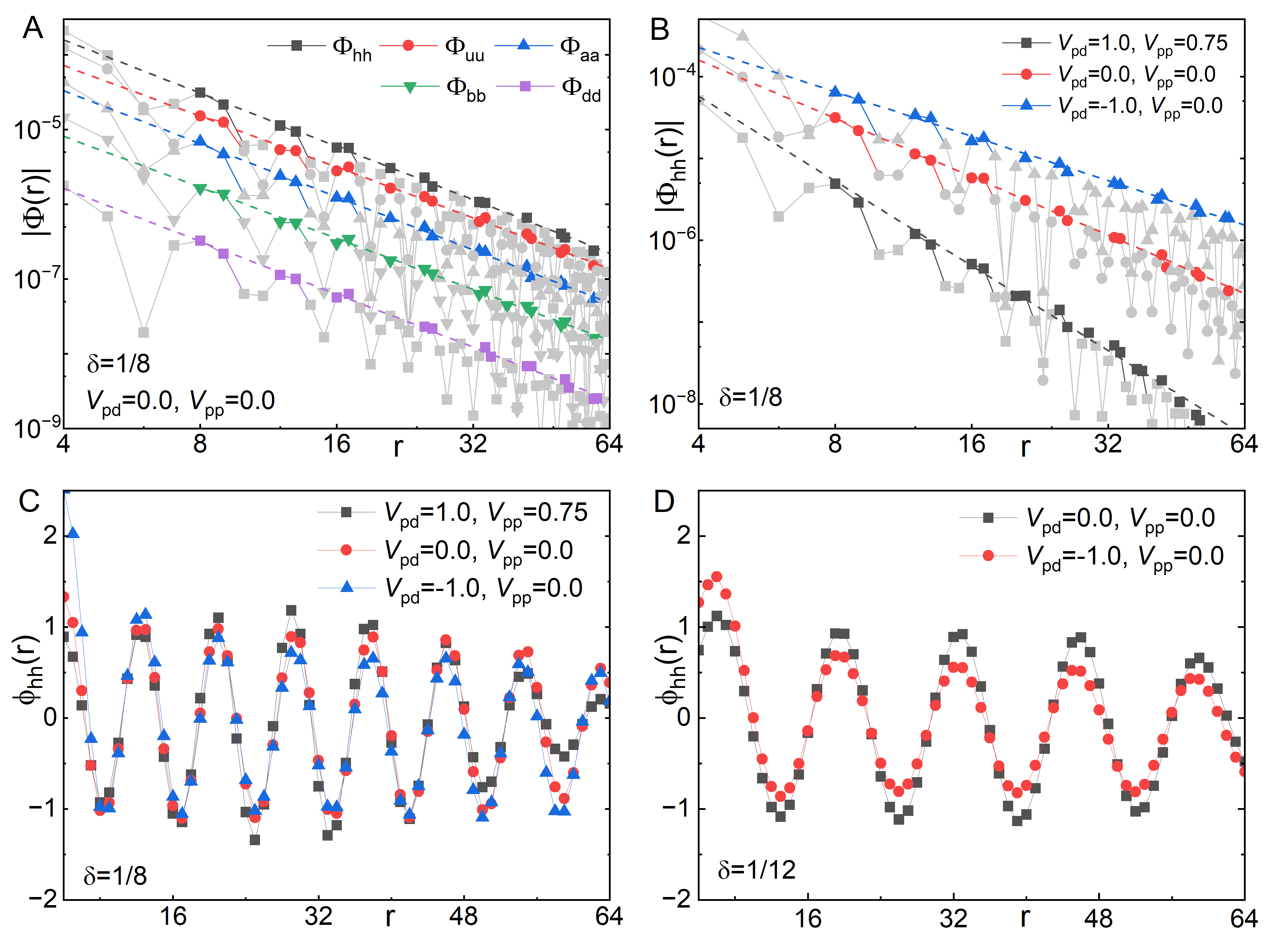}
\caption{Superconducting correlations. The magnitude of SC correlations $|\Phi_{\alpha\beta}(r)|$ are shown in (A) and $|\Phi_{hh}(r)|$ are shown in (B). The dashed lines represent fits to a power-law function $f(r) = A\ast r^{-K_{sc}}$. Data points away from the envelope and those at short distances are discarded in the fitting process and are shown in gray color. The normalized function $\phi_{hh}(r)=\Phi_{hh}(r)/f(r)$ reflect the spatial oscillation of $\Phi_{hh}(r)$ for $\delta=1/8$ in (C) and $\delta=1/12$ in (D).}\label{Fig:PDW}
\end{figure*}

\section{Model and Method}%
The lattice structure and the parameters of the three-band Hubbard model on the 2-leg square cylinder are shown in Fig.\ref{Fig:Lattice} where unit cell has a Cu $3d_{x^2-y^2}$ (Cu) orbital and oxygen $2p_x/2p_y$ (Ox/Oy) orbitals. Following Ref.\cite{White2015}, we consider the hole representation where the Hamiltonian is defined as%
\begin{eqnarray}
H &=& -t_{pd}\sum_{\langle ij\rangle\sigma}(\hat{d}_{i\sigma}^+\hat{p}_{j\sigma}+h.c.)-t_{pp}\sum_{\langle ij\rangle\sigma}(\hat{p}_{i\sigma}^+\hat{p}_{j\sigma}+h.c.)\nonumber \\%
&&+U_d\sum_i \hat{n}_{i\uparrow}^d \hat{n}_{i\downarrow}^d + U_p\sum_i \hat{n}_{i\uparrow}^p \hat{n}_{i\downarrow}^p + \Delta_{pd}\sum_{i\sigma}\hat{p}_{i\sigma}^+\hat{p}_{i\sigma} \nonumber\\%
&&+V_{pd}\sum_{\langle ij\rangle}\hat{n}_i^d \hat{n}_j^p + V_{pp}\sum_{\langle ij\rangle}\hat{n}_i^p \hat{n}_j^p.\label{Eq:Ham}
\end{eqnarray}
Here $\hat{d}_{i\sigma}^+$ and $\hat{p}_{j\sigma}^+$ create holes with spin-$\sigma$ on the $i^{th}$ Cu and $j^{th}$ oxygen sites, and $\langle ij\rangle$ denotes nearest-neighbor (NN) sites. $\hat{n}_{i\sigma}^d=\hat{d}_{i\sigma}^+ \hat{d}_{i\sigma}$ and $\hat{n}_{i\sigma}^p=\hat{p}_{i\sigma}^+ \hat{p}_{i\sigma}$ are number operators and $\hat{n}_i^{d/p}=\sum_\sigma \hat{n}_{i\sigma}^{d/p}$. $\Delta_{pd}=\epsilon_p-\epsilon_d$ is the energy difference between having a hole on the O site ($\epsilon_p$) versus a Cu site ($\epsilon_d$). $t_{pd}$ and $t_{pp}$ are the hole hopping matrix elements between NN Cu and O sites and NN O sites. $U_d$ and $U_p$ are the on-site Cu and O Coulomb repulsion, $V_{pd}$ and $V_{pp}$ are NN Cu-O and O-O interactions. Following \cite{White2015}, we have fixed the phases of orbitals such that the signs of hopping matrix elements remain the same throughout the lattice and are positive \footnote{This is equivalent to the usual three-band Hubbard model after a gauge transformation with the same ground state properties including correlations}. We set $t_{pd}=1$ as the energy unit and take a canonical set of parameters $t_{pp}=0.5$, $U_d=8$, $U_p=3$, $\Delta_{pd}=3$ for cuprates \cite{Armitage2010,Haule2014,White2015}. We study the ground state properties of the system as a function of $V_{pd}$ and $V_{pp}$.

In the present study, we take the lattice geometry to be cylindrical with periodic/open boundary condition in the $e_2$/$e_1$ direction, as shown in Fig.\ref{Fig:Lattice}. We focus on two-leg cylinders with width $L_y=2$ and length up to $L_x$=96, where $L_x$ and $L_y$ are the number of unit cells in the $e_1$ and $e_2$ directions, respectively. The total number of sites is $N=3L_x L_y+2L_y=3N_u+2L_y$, where $N_u$ is the number of unit cell. The overall hole density of the system is defined as $\rho=1+\delta$, where $\delta=N_h/N_u$ \footnote{Although $N\neq 3N_u$ such that the average value of $\delta$ differs slightly from $\bar{\delta}=3N_h/N$, deep in the bulk, i.e., relatively far from the open boundaries, $\delta=\bar{\delta}$} and $N_h$ denote the hole doping concentration and number of doped holes away from half-filling, respectively. We consider both $\delta=1/12$ and $\delta=1/8$ hole doping concentrations and perform up to 157 sweeps and keep up to $m=25000$ number of states with a typical truncation error $\epsilon\sim 10^{-9}$.

\section{Superconducting correlation}%
To test the possibility of superconductivity, we calculate the equal-time SC pair-field correlations. As the ground state with an even number of holes always have total spin 0, we focus on spin-singlet SC correlation \footnote{We have also calculated the spin-triplet SC correlations, which are much weaker than the spin-singlet SC correlations. This is expected as the ground state of the system is a spin-singlet state.}, which is defined as%
\begin{eqnarray}\label{Eq:SC_cor}
\Phi_{\alpha\beta}(r)=\langle\hat{\Delta}^{\dagger}_{\alpha}(x_0,y_0)\hat{\Delta}_{\beta}(x_0+r,y_0)\rangle.
\end{eqnarray}
Here, $\hat{\Delta}^{\dagger}_{\alpha}(x,y)=\frac{1}{\sqrt{2}}[\hat{c}^{\dagger}_{(x,y),\uparrow}\hat{c}^{\dagger}_{(x,y)+\alpha,\downarrow}-\hat{c}^{\dagger}_{(x,y),\downarrow}\hat{c}^{\dagger}_{(x,y)+\alpha,\uparrow}]$ is spin-singlet pair creation operator on the bond $\alpha=a$, $b$, $d$, $h$ and $u$ defined in Fig.\ref{Fig:Lattice}. ($x_0,y_0$) is a reference bond with $x_0\sim L_x/4$, $r$ is the distance between two bonds in the $e_1$ direction. We have calculated different components of the SC correlations, including $\Phi_{aa}$, $\Phi_{ab}$, $\Phi_{bb}$, $\Phi_{dd}(r)$, $\Phi_{d\bar{d}}(r)$, $\Phi_{\bar{d}\bar{d}}(r)$, $\Phi_{hh}$, $\Phi_{uu}$ and $\Phi_{uh}$ and find that $\Phi_{hh}(r)$ and $\Phi_{uu}(r)$ exhibit the strongest SC correlations as shown in Fig.\ref{Fig:PDW}A, i.e. the pairing is dominant between neighboring Cu sites. The pairing symmetry is consistent with $d$-wave, which is characterized by the fact $\Phi_{uu}(r)\sim \Phi_{hh}(r)\sim -\Phi_{uh}(r)$.

\begin{figure*}
  \includegraphics[width=0.9\linewidth]{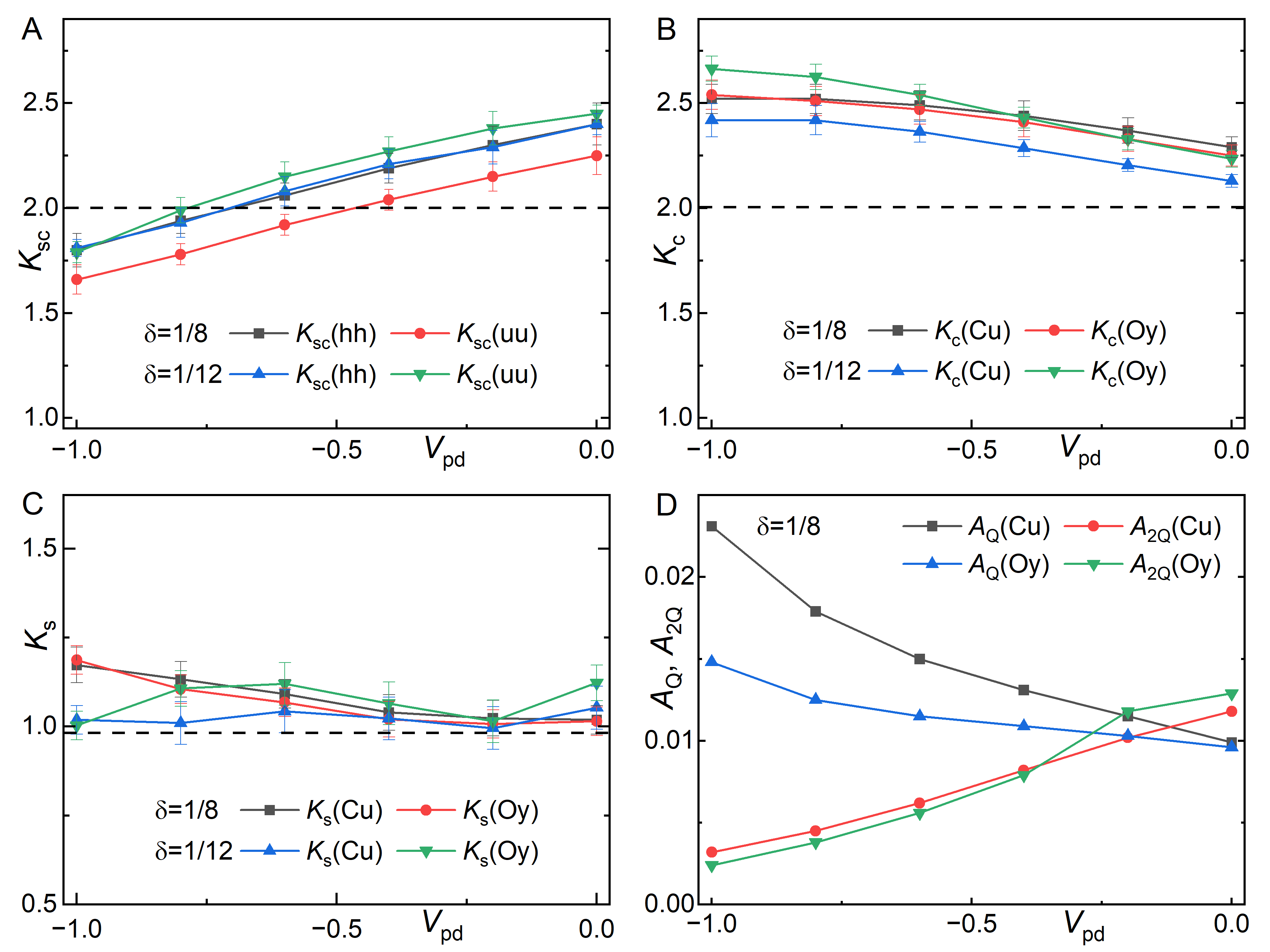}
  \caption{Luttinger exponents and CDW amplitudes. Extracted exponents $K_{sc}$ in (A), $K_c$ in (B), $K_s$ in (C), and CDW amplitudes $A_Q$ and $A_{2Q}$ in (D) as a function of $V_{pd}$ with $V_{pp}=0$. Error bars denote the numerical uncertainty and dash lines are guides for eyes.}\label{Fig:Exponent}
\end{figure*}

The spatial distribution of SC correlations $\Phi_{hh}(r)$ for the three representative choices of parameters are shown in  Fig.\ref{Fig:PDW}C-D: $\Phi_{hh}(r)$ exhibits clear spatial oscillations which can be well fitted by $\Phi_{hh}(r)\sim f(r)\phi_{hh}(r)$ for a large region of $r$, where $f(r)$ sets the envelope and $\phi_{hh}(r)$ determines the spatial oscillation. At long distances, the envelope function $f(r)$ is consistent with a power-law decay $f(r)=A\ast r^{-K_{sc}}$ with a Luttinger exponent $K_{sc}$. For instance, the extracted exponent $K_{sc}=3.4(2)$ at $V_{pd}=1.0$ and $V_{pp}=0.75$, $K_{sc}=2.4(1)$ at $V_{pd}=0$ and $V_{pp}=0$, and $K_{sc}=1.80(8)$ at $V_{pd}=-1.0$ even $V_{pp}=0$. It is worth emphasizing that the SC correlations can be notably enhanced by either reducing the NN repulsion or increasing the NN attraction, which is similar to the single-band Hubbard model \cite{Chen2021,Peng2022,Qu2021}. This is directly supported by the SC correlation in Fig.\ref{Fig:PDW}B and the value of $K_{sc}$ mentioned above. More complete results of $K_{sc}$ for $\Phi_{hh}$ and $\Phi_{uu}$ at $\delta=1/8$ and $\delta=1/12$ are shown in Fig.\ref{Fig:Exponent}A. This shows that the static PDW susceptibility diverges as $\chi_{pdw}\sim T^{-(2-K_{sc})}$ since $K_{sc}<2$ as $T\rightarrow 0$. As far as we know, to date, this is the strongest indication of PDW order that has been found in any DMRG study in doped Hubbard model on the square lattice.

The spatial oscillation of the SC correlations $\Phi(r)$ is characterized by the normalized function $\phi(r)=\Phi(r)/f(r)$ as mentioned above. Examples of $\phi_{hh}(r)$ for the representative choices of parameters are shown in Fig.\ref{Fig:PDW}C-D for $\delta=1/8$ and $\delta=1/12$, respectively. It is clear that $\phi_{hh}(r)$ oscillates periodically in real space and can be well fitted by $\phi_{hh}(r)\sim \sin (Qr +\theta)$, which is consistent with that of a PDW state \cite{Agterberg2020}. The PDW ordering wavevector $Q$ is incommensurate and hole doping dependent as $Q\approx 2\pi\delta$, corresponding to a wavelength $\lambda_{sc}\approx 1/\delta$, e.g., $\lambda_{sc}\approx 8$ at $\delta=1/8$ and $\lambda_{sc}\approx 12$ at $\delta=1/12$ as shown in Fig.\ref{Fig:PDW}C-D. This is distinct with $Q\approx 4\pi \delta$ in a qualitatively different system \cite{Xu2019}.

\begin{figure*}
  \includegraphics[width=0.9\linewidth]{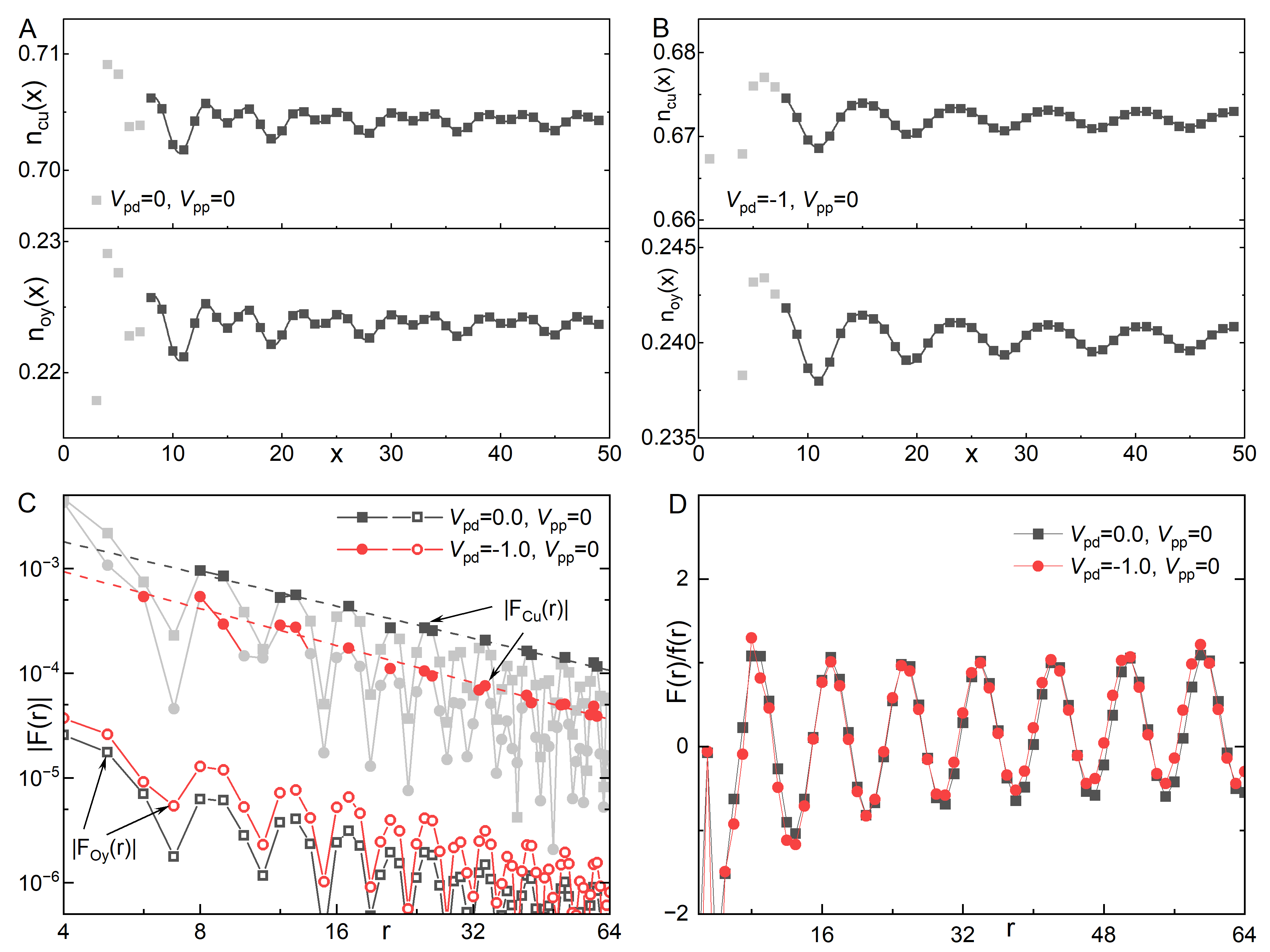}
\caption{Charge density profile and spin-spin correlations. Charge density profiles $n_{cu}(x)$ and $n_{oy}(x)$ in (A) for $V_{pd}=0$ and (B) for $V_{pd}=-1$. The solid lines denote fittings using Eq.(\ref{Eq:Kc}). The magnitude of spin-spin correlations $|F(r)|$ in (C) and its normalized function in (D). The dashed lines represent fits to a power-law function $f(r)=A\ast r^{-K_s}$. Here $\delta=1/8$ and data points away from the envelope and those at short distances are discarded in gray color in the fitting process.}\label{Fig:NxSS}
\end{figure*}

\section{Charge density wave order}%
To describe the charge density properties of the ground state of the system, we have calculated the charge density profile $n_\alpha(x,y)=\langle \hat{n}_\alpha(x,y)\rangle$ and its local rung average $n(x)={\sum_{y=1}^{L_y}} n_a(x,y)/L_y$, where $\alpha$=Cu/Ox/Oy site. Different from the single-band Hubbard model on square lattice \cite{White1997,Zheng2017,Jiang2019Hub,Jiang2020prr,Jiang2020prb}, the spatial oscillation of $n_\alpha(x)$ in the three-band Hubbard model is characterized by two wavevectors at $Q$ and $2Q$, corresponding to wavelengths $\lambda_Q\approx 1/\delta$ and $\lambda_{2Q}\approx 1/2\delta$, respectively. Examples of $n_{cu}(x)$ and $n_{oy}(x)$ for two representative sets of parameters are shown in Fig.\ref{Fig:NxSS}A-B.

At long distances, the spatial decay of the CDW correlation is dominated by a power-law with Luttinger exponent $K_c$, which can be obtained by fitting the charge density oscillations (Friedel oscillations) induced by the boundaries of the cylinder \cite{White2002}
\begin{eqnarray}\label{Eq:Kc}
n_\alpha(x)&=& A_Q\ast {\rm cos}(Qx + \phi_1)\ast x^{-K_c(\alpha)/2} \\ 
&+& A_{2Q}\ast {\rm cos}(2Qx + \phi_2)\ast x^{-K_c(\alpha)/2} + n_0(\alpha).\nonumber
\end{eqnarray}
Here $A_Q$ and $A_{2Q}$ are CDW amplitudes, $\phi_1$ and $\phi_2$ are phase shifts, $n_0$ is the mean density, and $Q\approx 2\pi\delta$. Similar with the single-band Hubbard model on four-leg square cylinders \cite{Peng2022}, we find that the CDW strength can be notably suppressed by NN attraction evidenced by the monotonic increase of $K_c$ with the increase of NN attraction $|V_{pd}|$ in Fig.\ref{Fig:Exponent}B. Examples of $K_c$ at $\delta=1/8$ and $V_{pp}=0$ are $K_c$(Cu)=2.29(5) and $K_c$(Oy)=2.25(5) for $V_{pd}$=0, and $K_c$(Cu)=2.52(7) and $K_c$(Oy)=2.54(7) for $V_{pd}$=-1. More complete results are shown in Fig.\ref{Fig:Exponent}B.

Interestingly, we find that the NN attraction can notably enhance the charge oscillation at $Q$ while suppressing the charge oscillation at $2Q$, evidenced by the CDW amplitudes $A_Q$ and $A_{2Q}$ as shown in Fig.\ref{Fig:Exponent}D. The charge oscillation is dominant at $2Q$ for strong NN repulsion, e.g., $A_{2Q}\approx 10A_{Q}$ at $V_{pd}=1.0$ and $V_{pp}=0.75$. When the NN interaction is weak, the charge oscillations at both $Q$ and $2Q$ become comparable with each other, e.g., $A_{2Q}\approx A_{Q}$ at $V_{pd}=V_{pp}=0$ in Fig.\ref{Fig:Exponent}D. However, when the NN interactions become attractive, e.g., $V_{pd}\leq -0.2$ in Fig.\ref{Fig:Exponent}D, the charge oscillation at $Q$ becomes dominant, e.g., $A_Q\gtrsim 6A_{2Q}$ at $V_{pd}=-1$ and $V_{pp}=0$.

\section{Spin-spin correlation}
To describe the magnetic properties of the ground state, we calculate the spin-spin correlation functions $F_\alpha(r)=\langle \vec{S}_{(x_0,y_0)}\cdot \vec{S}_{(x_0+r,y_0)}\rangle$, where $\alpha$=Cu/Ox/Oy site. Fig.\ref{Fig:NxSS}C show examples of $F_{Cu}(r)$ and $F_{Oy}(r)$ for two representative choices of parameters where we find that the spin-spin correlations between Cu sites are dominant evidenced by $|F_{Cu}(r)|\gg |F_{Oy}(r)|\sim |F_{Ox}(r)|$. Contrary to the exponential decaying spin-spin correlation in the Luther-Emery liquid of the square lattice single-band Hubbard model \cite{Jiang2018tJ,Jiang2019Hub,Jiang2020prr,Chung2020,Jiang2020prb}, we find that $F(r)$ in the lightly doped three-band Hubbard model decays as a power-law $F_\alpha(r)\sim r^{-K_s(\alpha)}$ at long distances, with a Luttinger exponent $K_s\sim 1$. For instance, $K_s$(Cu)=1.17(5) and $K_s$(Oy)=1.19(4) for $V_{pd}$=-1 and $V_{pp}$=0 at $\delta=1/8$. More complete results are shown in Fig.\ref{Fig:Exponent}C. Consistent with a PDW state, we find that $F(r)$ exhibits clear spatial oscillation as shown in Fig.\ref{Fig:NxSS}D at the PDW ordering wavevector $Q$ with a wavelength $\lambda_s \approx \lambda_{sc} \approx 1/\delta$, i.e., $\lambda_s\approx 8$ at $\delta=1/8$.

\begin{figure}
  \includegraphics[width=0.95\linewidth]{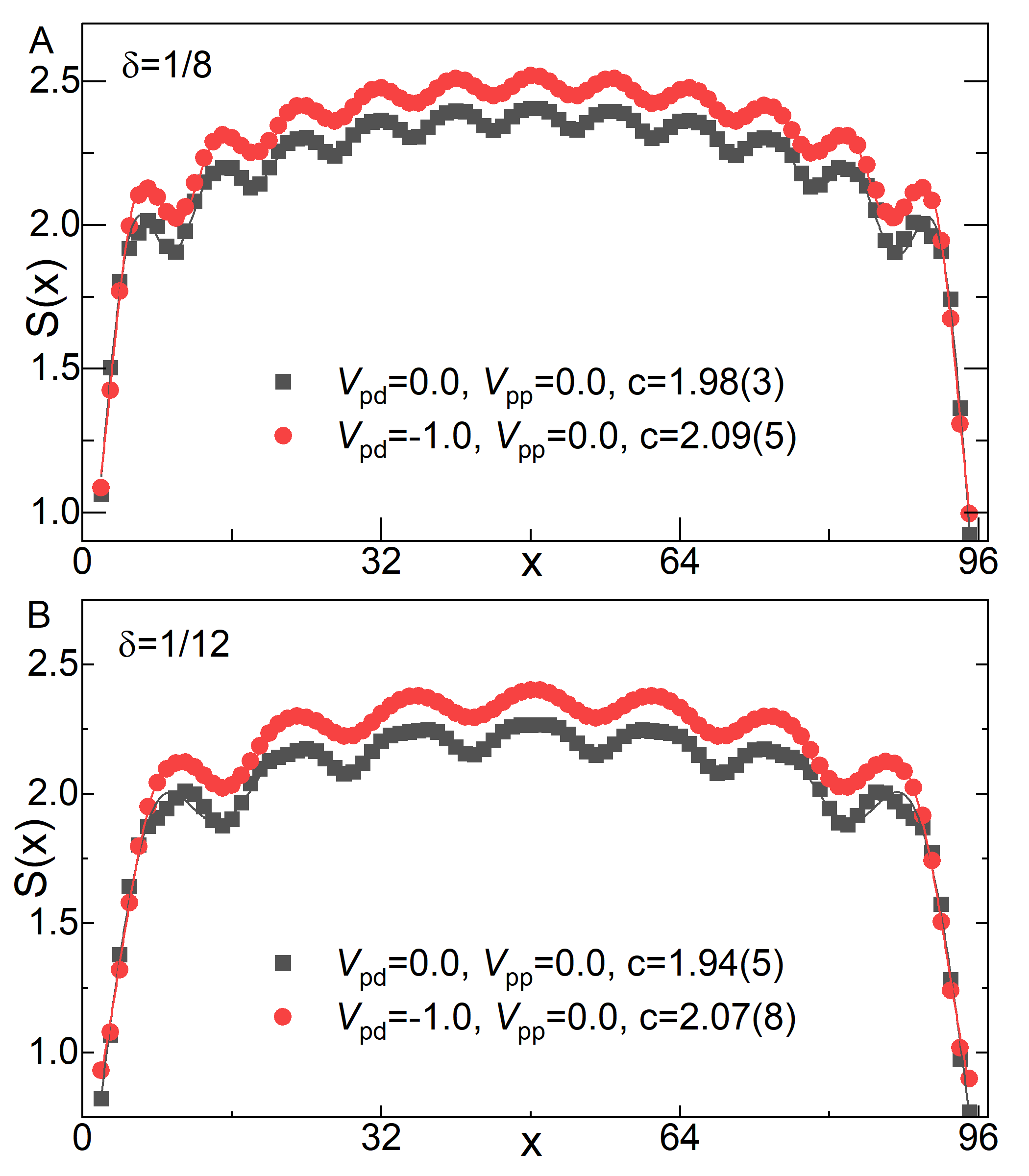}
\caption{Entanglement entropy. Von Neumann entanglement entropy $S(x)$ at $\delta=1/8$ in (A) and $\delta=1/12$ in (B). A couple of data points in gray color close to the open ends are excluded in the fitting to minimize the boundary effect.}\label{Fig:Entropy}
\end{figure}

\section{Entanglement entropy}%
Our results suggest that there are multiple gapless modes including charge and spin with a central charge $c$ which can be obtained by calculating the von Neumann entropy $S(x)=-{\rm Tr} \rho_x {\rm ln} \rho_x$, where $\rho_x$ is the reduced density matrix of a subsystem of length $x$. For critical systems in 1+1 dimensions described by a conformal field theory, it has been established\cite{Calabrese2004,Fagotti2011} that for an open system of length $L_x$, %
\begin{eqnarray}
S(x)&=&\frac{c}{6} \ln \big[\frac{4(L_x+1)}{\pi} \sin \frac{\pi(2x+1)}{2(L_x+1)}|\sin k_F|\big] \nonumber \\
&+&\tilde{A} \frac{\sin[k_F(2x+1)]}{\frac{4(L_x+1)}{\pi} \sin \frac{\pi(2x+1)}{2(L_x+1)}|\sin k_F|}+ \tilde{S},\label{Eq:EE}
\end{eqnarray}
where $\tilde{A}$ and $\tilde{S}$ are fitting parameters, and $k_F$ is the Fermi momentum. We find that the extracted $c\approx 2$ with examples shown in Fig.\ref{Fig:Entropy}. This suggests that there is one gapless charge mode and one gapless spin mode.

\section{Summary and Discussion}%
In summary, we have studied the ground state properties of hole-doped three-band Hubbard model on two-leg square cylinders with NN Cu-O and O-O interactions. Based on our results, we conclude that the ground state of the system is consistent with that of a PDW state, with mutually commensurate and power-law SC, CDW and SDW correlations. The Cooper pairing is dominant between neighboring Cu sites with $d$-wave symmetry. For modestly strong NN attraction, quasi-long-range PDW order with a divergent susceptibility emerges. As the critical attraction, where PDW quasi-long-range order develops, is close to that identified recently in cuprate Ba$_{2-x}$Sr$_x$CuO$_{3+\delta}$ \cite{Chen2021}, it may be reasonable to expect a comparable effective NN attraction in the CuO$_2$ plane given their chemical similarity.

When the NN attraction $V_{pd}\leq -0.2$ as shown in Fig.\ref{Fig:Exponent}D, it is clear that the CDW at wavevector $Q$ ($Q$-CDW) is dominant while the CDW at $2Q$ ($2Q$-CDW) is secondary. It is worth noting that our results suggest that the uniform SC component may be absent or at least small. If there was a small uniform component $\Delta_0$, then a $Q$-CDW could be generated by the product $\Delta_Q^P \Delta_0^\ast$ with PDW order $\Delta_Q^P$. Accordingly, the $2Q$-CDW could be generated by the product of PDW with itself $\Delta_Q^P \Delta_{-Q}^{P\ast}$. If this is the case, then the $Q$-CDW should be small as the uniform component is small, while the $2Q$-CDW should be dominant. However, this is contrary to our observations especially when $V_{pd}\leq -0.2$. Alternatively, the $2Q$-CDW could come from a composite of two SDW at $Q$, which is, however, inconsistent with our observations. This is because the $2Q$-CDW weakens notably with the increase of NN attraction when $V_{pd}\leq -0.2$, while the SDW remains unchanged. Although the favored forms of order, with SC and SDW at the same $Q$, and CDW at both $Q$ and $2Q$, are reminiscent of those conjectured to be present in the cuprates BSCCO and LBCO \cite{Edkins2019,Liu2021,Li2007,Tranquada2020,Tranquada2021}, it is worth emphasizing that the underlying mechanism of our observations that all three orders (PDW, CDW and SDW) are all at the same $Q$ can be very different from the stripe picture of PDW \cite{Agterberg2020}.

We are grateful to Steven Kivelson, Thomas Devereaux, John Tranquada, Patrick Lee and Ziqiang Wang for insightful discussions and invaluable suggestions, and Patrick Lee for pointing out the importance of three orders at the same $Q$. This work was supported by the Department of Energy, Office of Science, Basic Energy Sciences, Materials Sciences and Engineering Division, under Contract DE-AC02-76SF00515.

%

\end{document}